\documentstyle[12pt]{article}
\newenvironment{acknowledgement}%
  {\bigskip\noindent{\large\bf Acknowledgement}\medskip\par\noindent}%
  {\bigskip}
\newenvironment{note-added}%
  {\bigskip\noindent{\large\bf Note added}\medskip\par\noindent}%
  {\bigskip}
\newcommand{\beqn}{\begin{equation}}
\newcommand{\eeqn}{\end{equation}}
\newcommand{\beqnarray}{\begin{eqnarray}}
\newcommand{\eeqnarray}{\end{eqnarray}}
\newcommand{\rd}{\partial}
\newcommand{\dfrac}[2]{ \frac{\displaystyle #1}{\displaystyle #2} }

\newcommand{\res}{\;\mathop{\mbox{\rm res}}}
\newcommand{\cA}{{\cal A}}
\newcommand{\cAbar}{\bar{\cA}}
\newcommand{\cB}{{\cal B}}
\newcommand{\cBbar}{\bar{\cB}}
\newcommand{\cL}{{\cal L}}
\newcommand{\cLbar}{\bar{\cL}}
\newcommand{\cM}{{\cal M}}
\newcommand{\cMbar}{\bar{\cM}}
\newcommand{\cS}{{\cal S}}
\newcommand{\cSbar}{\bar{\cS}}
\newcommand{\cR}{{\cal R}}
\newcommand{\cO}{{\cal O}}

\newcommand{\tbar}{\bar{t}}
\newcommand{\ubar}{\bar{u}}
\newcommand{\vbar}{\bar{v}}
\newcommand{\Nbar}{\bar{N}}
\newcommand{\Wbar}{\bar{W}}
\newcommand{\lambdabar}{\bar{\lambda}}
\newcommand{\mubar}{\bar{\mu}}
\newcommand{\phibar}{\bar{\phi}}
\newcommand{\Phibar}{\bar{\Phi}}
\newcommand{\Psibar}{\bar{\Psi}}
\newcommand{\dbra}{\left<\!\left<}
\newcommand{\dket}{\right>\!\right>}
\begin{document}

\begin{flushright}
\baselineskip=12pt
KUCP-0067\\
hep-th/9403190\\
March 1994
\end{flushright}

\begin{center}
\LARGE
    Dispersionless Toda Hierarchy and \\
    Two-dimensional String Theory\\
\bigskip
\Large
    Kanehisa Takasaki \\
\normalsize\it
    Department of Fundamental Sciences\\
    Faculty of Integrated Human Studies, Kyoto University\\
    Yoshida-Nihonmatsu-cho, Sakyo-ku, Kyoto 606, Japan\\
\rm
    E-mail: takasaki @ jpnyitp (Bitnet)\\
\end{center}
\bigskip
\begin{abstract}

\noindent
The dispersionless Toda hierarchy turns out to lie in the
heart of a recently proposed Landau-Ginzburg formulation of
two-dimensional string theory at self-dual compactification
radius. The dynamics of massless tachyons with discrete
momenta is shown to be encoded into the structure of a
special solution of this integrable hierarchy. This solution
is obtained by solving a Riemann-Hilbert problem. Equivalence
to the tachyon dynamics is proven by deriving recursion
relations of tachyon correlation functions in the machinery
of the dispersionless Toda hierarchy. Fundamental ingredients
of the Landau-Ginzburg formulation, such as Landau-Ginzburg
potentials and tachyon Landau-Ginzburg fields, are translated
into the language of the Lax formalism. Furthermore, a wedge
algebra is pointed out to exist behind the Riemann-Hilbert
problem, and speculations on its possible role as generators
of ``extra'' states and fields are presented.

\end{abstract}
\newpage

\section{Introduction}

Recently a Landau-Ginzburg model of two-dimensional
strings at self-dual radius (i.e., $c = 1$ topological
matter coupled to two-dimensional gravity) has been
proposed and studied by several groups
\cite{bib:Ghoshal-Mukhi,bib:HOP,bib:Danielsson}.
This model is in a sense a natural extrapolation of the
topological $A_{k+1}$ model to $k = -3$, and seems to
inherits the remarkable properties of the $A_{k+1}$ models
such as:
(i) an underlying structure of Lax equation \cite{bib:DVV},
(ii) a period integral representation of correlation
functions \cite{bib:period-integral},
(iii) an algebraic structure of gravitational primaries
and descendents \cite{bib:contact-algebra}, etc.
Although the status of the so called special (discrete)
states \cite{bib:special-state} still remains obscure,
the dynamics of massless tachyons with discrete momenta
is shown to be correctly described in this new framework.

The $c = 1$ model, however, differs from the $A_{k+1}$
(and some other $c < 1$) models in several essential
aspects.  This seems to be eventually due to the difference
of underlying integrable hierarchies. The $A_{k+1}$ models
are special solutions of the dispersionless KP (or
generalized KdV) hierarchy
\cite{bib:Krichever,bib:Dubrovin,bib:TT-dKP}. Hanany et al.
\cite{bib:HOP} suggested a similar link between the $c = 1$
model and the dispersionless Toda hierarchy \cite{bib:TT-dToda}.

In this paper, we demonstrate the suggestion of Hanany et al.
in the machinery of dispersionless Toda hierarchy, and search
for implications therefrom. Our basic observation is that the
tachyon dynamics at self-dual radius is perfectly encoded into
the structure of a special solution of this integrable hierarchy.
In Section \ref{sec:notions}, we recall fundamental notions
concerning the dispersionless Toda hierarchy, and in Sections
\ref{sec:lambda}-\ref{sec:symmetries}, reformulate several
results of our previous work \cite{bib:TT-dToda} in a more
convenient form. The aforementioned special solution is
constructed in Section \ref{sec:RHproblem} by solving a
Riemann-Hilbert problem. A set of $w_{1+\infty}$-constraints
(recursion relations) characterizing tachyon correlation
functions are derived from the Riemann-Hilbert problem in
Section \ref{sec:constraints}. Since, as remarked by Hanany et al.,
those $w_{1+\infty}$-constraints determine tachyon correlation
functions uniquely, we can conclude that our solution indeed
describe the tachyon dynamics. In Section \ref{sec:wedge} we
show the existence of a wedge algebra behind the Riemann-Hilbert
problem, and propose a speculative interpretation of this algebra
as generators of ``extra'' states and fields in the $c = 1$ model.
Section \ref{sec:conclusion} is devoted to concluding remarks.

\section{Fundamental notions in dispersionless Toda hierarchy
\label{sec:notions}}

The Lax formalism of the dispersionless Toda hierarchy
is based on the two-dimensional Poisson bracket
\beqn
    \{ A(p,s), B(p,s) \}
    =  p \frac{\rd A}{\rd p} \frac{\rd B}{\rd s}
     - \frac{\rd A}{\rd s} p \frac{\rd B}{\rd p}
\eeqn
rather than usual commutators. Fundamental quantities
(counterparts of Lax operators) are two Laurent series
$\cL$ and $\cLbar$ of the form
\beqnarray
  \cL &=& p + \sum_{n=0}^\infty u_{n+1}(t,\tbar,s) p^{-n}, \\
  \cLbar^{-1} &=& \ubar_0 p^{-1}
                  + \sum_{n=0}^\infty \ubar_{n+1}(t,\tbar,s) p^n,
                                      \label{eq:L}
\eeqnarray
where the coefficients depend on time variables of flows
$t = (t_1,t_2,\ldots)$ and $\tbar = (\tbar_1,\tbar_2,\ldots)$
as well as the spatial coordinate $s$. (We have slightly
changed notations in the previous work \cite{bib:TT-dToda}.)
Lax equations of these ``Lax functions" are written
\beqnarray
    \dfrac{\rd \cL}{\rd t_n} = \{ \cB_n, \cL \},        &\quad&
    \dfrac{\rd \cL}{\rd \tbar_n} = \{ \cBbar_n, \cL \}, \nonumber \\
    \dfrac{\rd \cLbar}{\rd t_n} = \{ \cB_n, \cLbar \},  &\quad&
    \dfrac{\rd \cLbar}{\rd \tbar_n} = \{ \cBbar_n, \cLbar \},
                                        \label{eq:LaxL}
\eeqnarray
where $\cB_n$ and $\cBbar_n$ are given by
\beqnarray
  &&  \cB_n = ( \cL^n )_{\ge 0},  \quad
      \cBbar_n = ( \cLbar^{-n} )_{\le -1},                      \\
  && (\quad)_{\ge 0}:\
        \mbox{projection onto}\ p^0,p^1,\ldots,        \nonumber\\
  && (\quad)_{\le -1}:\
        \mbox{projection onto}\ p^{-1},p^{-2},\ldots.  \nonumber
\eeqnarray
Furthermore, given such a pair $\cL$ and $\cLbar$, one can find
another pair of Laurent series
\beqnarray
    \cM &=&  \sum_{n=1}^\infty n t_n \cL^n + s
           + \sum_{n=1}^\infty v_n(t,\tbar,s)\cL^{-n}, \nonumber\\
    \cMbar &=& - \sum_{n=1}^\infty n \tbar_n \cLbar^{-n} + s
               + \sum_{n=1}^\infty \vbar_n(t,\tbar,s) \cLbar^n
                                               \label{eq:M}
\eeqnarray
that satisfy the Lax equations
\beqnarray
    \dfrac{\rd \cM}{\rd t_n} = \{ \cB_n, \cM \},        &&
    \dfrac{\rd \cM}{\rd \tbar_n} = \{ \cBbar_n, \cM \}, \nonumber \\
    \dfrac{\rd \cMbar}{\rd t_n} = \{ \cB_n, \cMbar \},  &&
    \dfrac{\rd \cMbar}{\rd \tbar_n} = \{ \cBbar_n, \cMbar \}
                                               \label{eq:LaxM}
\eeqnarray
and the canonical Poisson relations
\beqn
    \{ \cL, \cM \} = \cL, \quad
    \{ \cLbar,\cMbar \} = \cLbar.         \label{eq:canonical}
\eeqn
It is rather these ``extra'' Lax functions that play a
central role in our approach to two-dimensional strings.

Before going forward, a few comments on formal residue
calculus are in order. We consider residues as being
defined for 1-forms as:
\beqn
    \res \sum a_n z^n dz = a_{-1}.
\eeqn
Residues of more general 1-form are to be evaluated
by the standard rule of exterior differential calculus:
\beqn
    \res f(z)dg(z) = \res f(z) \frac{dg(z)}{dz} dz.
\eeqn
Residues thus defined are invariant under coordinate
transformations  $z \to w = h(z)$ sending $\infty\to\infty$
or $0\to0$.

We can now define four fundamental potentials $\phi$,
$F$, $\cS$ and $\cSbar$ as follows.

The first potential $\phi = \phi(t,\tbar,s)$ is defined
by the equation
\beqn
    d\phi =  \sum_{n=1}^\infty \res( \cL^n d\log p) dt_n
      -\sum_{n=1}^\infty \res( \cLbar^{-n} d\log p) d\tbar_n
      + \log \ubar_0 ds,
                                            \label{eq:dphi}
\eeqn
where ``$d$" means total differentiation in $(t,\tbar,s)$,
and of course $d\log p = dp/p$.  The right hand side is a
closed form as far as $\cL$ and $\cLbar$ are subject to
Lax equations (\ref{eq:LaxL}). This potential $\phi$
satisfies the second-order equation
\beqn
    \dfrac{\rd^2 \phi}{\rd t_1 \rd \tbar_1}
    + \frac{\rd}{\rd s} \exp\left( \frac{\rd\phi}{\rd s} \right)
    = 0.
\eeqn
This is the well known dispersionless (or long-wave) limit of
the two-dimensional Toda field equation.

The second potential $F = F(t,\tbar,s)$ is defined by
the equation
\beqn
    dF = \sum_{n=1}^\infty v_n dt_n
         - \sum_{n=1}^\infty \vbar_n d\tbar_n
         + \phi ds.
                                                \label{eq:dF}
\eeqn
Again, the right hand side is a closed form as far as
$\cL$, $\cM$, $\cLbar$ and $\cMbar$ are subject to
Lax equations (\ref{eq:LaxL},\ref{eq:LaxM},\ref{eq:canonical}).
This potential $F$ plays the role of a ``generating
function" --- all other quantities $u_n$, $\ubar_n$, $v_n$,
$\vbar_n$ and $\phi$ can be reproduced from $F$ by
differentiation with respect to $t$, $\tbar$ and $s$.
This is obviously reminiscent of the role of partition
functions with external sources in usual field theories.
In our earlier work \cite[1991]{bib:TT-dToda},
$F$ was defined as logarithm of the ``tau function" of
the dispersionless Toda hierarchy, but it was later
recognized that $F$ is also connected with the tau
function $\tau(\hbar,t,\tbar,s)$ of the full Toda
hierarchy by $\hbar$-expansion \cite[1993]{bib:TT-dToda}:
\beqn
    \log\tau(\hbar,t,\tbar,s)
    = \hbar^{-2} F(t,\tbar,s) + O(\hbar^{-1}).
\eeqn

The last two potentials $\cS = \cS(t,\tbar,s,p)$ and
$\cSbar = \cSbar(t,\tbar,s,p)$ can be defined rather directly as:
\beqnarray
    \cS &=&
      \sum_{n=1}^\infty t_n \cL^n + s \log\cL
      - \sum_{n=1}^\infty
          \frac{1}{n} \dfrac{\rd F}{\rd t_n} \cL^{-n},
                                                     \nonumber \\
    \cSbar &=&
      \sum_{n=1}^\infty \tbar_n \cLbar^{-n} + s\log\cLbar + \phi
      - \sum_{n=1}^\infty
          \frac{1}{n} \dfrac{\rd F}{\rd \tbar_n} \cLbar^n.
                                                \label{eq:S}
\eeqnarray
We call $\cS$ and $\cSbar$  ``potentials" because they can also
be characterized as:
\beqnarray
    d\cS &=& \cM d\log\cL + \log p ds
             + \sum_{n=1}^\infty \cB_n dt_n
             + \sum_{n=1}^\infty \cBbar_n d\tbar_n,
                                                    \nonumber \\
    d\cSbar &=& \cMbar d\log\cLbar + \log p ds
                + \sum_{n=1}^\infty \cB_n dt_n
                + \sum_{n=1}^\infty \cBbar_n d\tbar_n,
                                                   \label{eq:dS}
\eeqnarray
where ``$d$" now means total differentiation in $(t,\tbar,s)$
and $p$. An immediate consequence of (\ref{eq:dS}) is the
following expressions of $\cB_n$ and $\cBbar_n$:
\beqnarray
    \cB_m
      &=& \cL^m
          - \sum_{n=1}^\infty \frac{1}{n}
               \dfrac{\rd^2 F}{\rd t_m \rd t_n}
                   \cL^{-n}                           \nonumber \\
      &=& \dfrac{\rd^2 F}{\rd t_m \rd s}
          - \sum_{n=1}^\infty \frac{1}{n}
               \dfrac{\rd^2 F}{\rd t_m \rd \tbar_n}
                   \cLbar^n,                          \nonumber \\
    \cBbar_m
      &=& - \sum_{n=1}^\infty \frac{1}{n}
               \dfrac{\rd^2 F}{\rd \tbar_m \rd t_n}
                   \cL^{-n}                           \nonumber \\
      &=& \cLbar^{-m}
          + \dfrac{\rd^2 F}{\rd \tbar_m \rd s}
          - \sum_{n=1}^\infty \frac{1}{n}
               \dfrac{\rd^2 F}{\rd \tbar_m \rd \tbar_n}
                   \cLbar^n.
                                                   \label{eq:BbyL}
\eeqnarray

\section{Spectral parameters $\lambda$ and $\lambdabar$
\label{sec:lambda}}

We now introduce two new variables $\lambda$ and $\lambdabar$,
and reformulate the setting of the previous section by
replacing
\beqn
    \cL \to \lambda, \quad
    \cLbar \to \lambdabar.           \label{eq:L->lambda}
\eeqn

By this substitution, $\cS$ and $\cSbar$ are replaced by
\beqnarray
    \cS(\cL\to\lambda)
    &=& \sum_{n=1}^\infty t_n \lambda^n
        + s \log\lambda
        - \sum_{n=1}^\infty
            \frac{1}{n} \dfrac{\rd F}{\rd t_n} \lambda^{-n},
                                                    \nonumber \\
    \cSbar(\cLbar\to\lambdabar)
    &=& \sum_{n=1}^\infty \tbar_n \lambdabar^{-n}
        + s\log\lambdabar + \phi
        - \sum_{n=1}^\infty
            \frac{1}{n} \dfrac{\rd F}{\rd \tbar_n} \lambdabar^n.
                                            \label{eq:S(lambda)}
\eeqnarray
In the language of the full Toda hierarchy, these quantities
are just the leading terms in $\hbar$-expansion of logarithm
of two Baker-Akhiezer functions $\Psi(\hbar,t,\tbar,s,\lambda)$
and $\Psibar(\hbar,t,\tbar,s,\lambdabar)$ \cite{bib:TT-dToda}:
\beqnarray
    \Psi(\hbar,t,\tbar,s,\lambda)
      &=& \exp[ \hbar^{-1} \cS(\cL\to\lambda)
          + O(\hbar^0)],
                                               \nonumber \\
    \Psibar(\hbar,t,\tbar,s,\lambdabar)
      &=& \exp[ \hbar^{-1} \cSbar(\cLbar\to\lambdabar)
          + O(\hbar^0)].
\eeqnarray
The new variables $\lambda$ and $\lambdabar$ are thus nothing
but the spectral parameters of the full Toda hierarchy.
In the usual setting, actually, one does not have to
distinguish between $\lambda$ and $\lambdabar$; in the
present setting, they correspond to the two Lax functions
$\cL$ and $\cLbar$. Furthermore, in our interpretation of
the Laudau-Ginzburg formulation, they do arise in a
different form as we shall see later. These are main
reasons that we use the two different spectral parameters.

Similarly, $\cM$ and $\cMbar$ are replaced by
\beqnarray
  \cM(\cL\to\lambda)
     &=& \sum_{n=1}^\infty n t_n \lambda^n + s
         + \sum_{n=1}^\infty \dfrac{\rd F}{\rd t_n} \lambda^{-n},
                                                     \nonumber \\
  \cMbar(\cLbar\to\lambdabar)
     &=& - \sum_{n=1}^\infty n \tbar_n \lambdabar^{-n} + s
         - \sum_{n=1}^\infty
             \dfrac{\rd F}{\rd \tbar_n} \lambdabar^n,
                                           \label{eq:M(lambda)}
\eeqnarray
where we have rewritten $v_n$ and $\vbar_n$ into derivatives
of $F$. By comparing (\ref{eq:M(lambda)}) with (\ref{eq:S(lambda)}),
one can readily find that
\beqnarray
    \cM(\cL\to\lambda)
        &=& \lambda\dfrac{\rd}{\rd\lambda} \cS(\cL\to\lambda),
                                                   \nonumber \\
    \cMbar(\cLbar\to\lambdabar)
        &=& \lambdabar\dfrac{\rd}{\rd\lambdabar}
            \cSbar(\cLbar\to\lambdabar).
                                 \label{eq:M(lambda)byS(lambda)}
\eeqnarray
These equations can be derived from (\ref{eq:dS}), too.

Lastly, applying the same substitution rule to (\ref{eq:BbyL}),
we can define four quantities $\cB_n(\cL\to\lambda)$,
$\cB_n(\cLbar\to\lambdabar)$, $\cBbar_n(\cL\to\lambda)$,
$\cBbar_n(\cLbar\to\lambdabar)$. Eqs. (\ref{eq:dS}) imply
that these quantities, too, can be written as derivatives
of $\cS(\cL\to\lambda)$ and $\cSbar(\cLbar\to\lambdabar)$:
\beqnarray
    \cB_n(\cL\to\lambda)
        = \dfrac{\rd}{\rd t_n} \cS(\lambda),       &&
    \cBbar_n(\cL\to\lambda)
        = \dfrac{\rd}{\rd \tbar_n} \cS(\lambda),   \nonumber\\
    \cB_n(\cLbar\to\lambdabar)
        = \dfrac{\rd}{\rd t_n} \cSbar(\lambdabar), &&
    \cBbar_n(\cLbar\to\lambdabar)
        = \dfrac{\rd}{\rd \tbar_n} \cSbar(\lambdabar).
                               \label{eq:B(lambda)byS(lambda)}
\eeqnarray

An immediate consequence of (\ref{eq:M(lambda)byS(lambda)})
and (\ref{eq:B(lambda)byS(lambda)}) is the following identities:
\beqnarray
    \dfrac{\rd}{\rd\lambda} \cB_n(\cL\to\lambda)
     &=& \dfrac{\rd}{\rd t_n}
            \cM(\cL\to\lambda) \lambda^{-1},
                                                \nonumber \\
    \dfrac{\rd}{\rd\lambda} \cBbar_n(\cL\to\lambda)
     &=& \dfrac{\rd}{\rd \tbar_n}
            \cM(\cL\to\lambda) \lambda^{-1},
                                                \nonumber \\
    \dfrac{\rd}{\rd (\lambdabar^{-1})}
       \cB_n(\cLbar\to\lambdabar)
     &=& - \dfrac{\rd}{\rd t_n}
              \cMbar(\cLbar\to\lambdabar) \lambdabar,
                                                \nonumber \\
    \dfrac{\rd}{\rd (\lambdabar^{-1})}
       \cBbar_n(\cLbar\to\lambdabar)
     &=& - \dfrac{\rd}{\rd \tbar_n}
              \cMbar(\cLbar\to\lambdabar) \lambdabar.
                                  \label{eq:dB(lambda)dM(lambda)}
\eeqnarray
We shall show later that these quantities are fundamental
ingredients of the Landau-Ginzburg formulation of
two-dimensional strings.

\section{Symmetries of dispersionless Toda hierarchy
\label{sec:symmetries}}

Given two functions $\cA = \cA(\cL,\cM)$ and $\cAbar =
\cAbar(\cLbar,\cMbar)$, one can construct an infinitesimal
symmetry $\delta_{\cA,\cAbar}$ of the dispersionless
Toda hierarchy \cite{bib:TT-dToda}.  More precisely,
$\cA$ and $\cAbar$ are assumed to be a ``good'' function,
such as a polynomial of $(\cL,\cL^{-1},\cM)$ and
$(\cLbar,\cLbar^{-1},\cMbar)$, respectively, with
constant coefficients. We here explain how these symmetries
are actually defined, and present several formulas that
we shall use crucially in the subsequent sections.

Let us consider the ring $\cR$ generated by $t$, $\tbar$,
$s$, $F$ and all its derivatives. In this setting, $F$ and
its derivatives have to be considered abstract ``symbols''
rather than actual functions of $(t,\tbar,s)$. By
``derivation'' we mean a linear map $\delta: \cR \to \cR$
satisfying the Leibniz rule
$\delta(ab) = \delta(a) b + a \delta(b)$. One can define
the derivations $\rd/\rd t_n$, $\rd/\rd\tbar_n$, and
$\rd/\rd s$ as derivations on $\cR$ in an obvious manner:
\beqnarray
  && \dfrac{\rd}{\rd t_n} F = v_n, \quad
     \dfrac{\rd}{\rd \tbar_n} F = - \vbar_n, \quad
     \dfrac{\rd}{\rd s} F = \phi,             \nonumber\\
  && \dfrac{\rd}{\rd t_n} t_m = \delta_{nm}, \quad
     \ldots \mbox{etc} \ldots.
\eeqnarray
Differential equations satisfied by $F$ and its derivatives
(which include differential equations of $v_n$, $\vbar_n$
and $\phi$, too) are thus encoded into these
differential-algebraic structures of $\cR$.

The symmetry $\delta_{\cA,\cAbar}$ is defined to be an
additional derivation of $\cR$ with the following properties
\cite{bib:TT-dToda}:
\begin{itemize}
    \item The action of $\delta_{\cA,\cAbar}$ on $F$ is given by
\beqn
    \delta_{\cA,\cAbar} F
    = - \res\left( \int_0^{\cM(\cL\to\lambda)}
            \cA(\lambda,\mu)d\mu \right) d\lambda
      + \res\left( \int_0^{\cMbar(\cLbar\to\lambdabar)}
            \cAbar(\lambdabar,\mubar)d\mubar \right) d\lambdabar.
                                         \label{eq:deltaF}
\eeqn
    \item $\delta_{\cA,\cAbar}$ acts trivially on $t$, $\tbar$
    and $s$ as:
\beqn
    \delta_{\cA,\cAbar} t_n = 0, \quad
    \delta_{\cA,\cAbar} \tbar_n = 0, \quad
    \delta_{\cA,\cAbar} s = 0.
\eeqn
    \item $\delta_{\cA,\cAbar}$ commutes with $\rd/\rd t_n$,
    $\rd/\rd \tbar_n$ and $\rd/\rd s$:
\beqn
    \left[ \delta_{\cA,\cAbar}, \frac{\rd}{\rd t_n} \right] =
    \left[ \delta_{\cA,\cAbar}, \frac{\rd}{\rd \tbar_n} \right] =
    \left[ \delta_{\cA,\cAbar}, \frac{\rd}{\rd s} \right] = 0.
\eeqn
\end{itemize}
The last property implies, in particular, that
$\delta_{\cA,\cAbar}$ commutes with all flows of the
dispersionless Toda hierarchy, a condition characterizing
a symmetry!

Furthermore, these symmetries satisfy the following commutation
relations \cite{bib:TT-dToda}:
\beqn
    [\delta_{\cA,\cAbar}, \delta_{\cB,\cBbar} ]
    = \delta_{\{\cA,\cB\},\{\cAbar,\cBbar\}}
      + \res\left( \cA(\lambda,0)d\cB(\lambda,0)
                  -\cAbar(\lambdabar,0)d\cBbar(\lambdabar,0)
        \right) \rd_F,
                                              \label{eq:CR}
\eeqn
where $\rd_F$ is yet another derivation on $\cR$ defined by
\beqn
    \rd_F F = 0, \quad
    \rd_F (\mbox{any other generator of}\ \cR) = 0,
\eeqn
which accordingly commute with all other derivations $\rd/\rd t_n$,
$\rd/\rd \tbar_n$, $\rd/\rd s$ and $\delta_{\cA,\cAbar}$. Thus an
underlying Lie algebra is a central extension of $w_{1+\infty}
\oplus w_{1+\infty}$; note that $w_{1+\infty}$ is now realized
as the Lie algebra of Poisson brackets.

The action of $\delta_{\cA,\cAbar}$ on other fundamental
quantities such as $v_n$, $\vbar_n$ and $\phi$, etc. can
be read off from the above construction, because they all
are derivatives of $F$. For $v_n$, $\vbar_n$ and $\phi$,
we have the following formulas (and, actually, the above
formula for $F$ was first discovered by ``integrating''
these formulas \cite{bib:TT-dToda}):
\beqnarray
    \delta_{\cA,\cAbar} v_n
      &=& \res \left( - \cA(\cL,\cM)
            + \cAbar(\cLbar,\cMbar) \right) d\cB_n,    \nonumber \\
    \delta_{\cA,\cAbar} \vbar_n
      &=& \res \left( + \cA(\cL,\cM)
            - \cAbar(\cLbar,\cMbar) \right) d\cBbar_n, \nonumber \\
    \delta_{\cA,\cAbar} \phi
      &=& \res \left( - \cA(\cL,\cM)
            + \cAbar(\cLbar,\cMbar) \right) d\log p.
                                             \label{eq:deltav}
\eeqnarray
Furthermore, since $\cM(\cL\to\lambda)$ and
$\cMbar(\cLbar\to\lambdabar)$ are generating functions of
$v_n$ and $\vbar_n$, one should be able to rewrite the first two
of (\ref{eq:deltav}) in terms of these generating functions.
This indeed results in the following formulas:
\beqnarray
    \delta_{\cA,\cAbar} \cM(\cL\to\lambda)
      &=& \lambda\frac{\rd}{\rd\lambda}
           \left[ \left(
                 \cA(\cL,\cM) - \cAbar(\cLbar,\cMbar)
           \right)_{\le -1}(\cL \to \lambda) \right] ,
                                                  \nonumber \\
    \delta_{\cA,\cAbar} \cMbar(\cLbar\to\lambdabar)
      &=& \lambdabar\frac{\rd}{\rd\lambdabar}
           \left[ \left(
                 - \cA(\cL,\cM) + \cAbar(\cLbar,\cMbar)
           \right)_{\ge 0}(\cLbar \to \lambdabar) \right] ,
                                  \label{eq:deltaM(lambda)}
\eeqnarray
where $\delta_{\cA,\cAbar}$ is understood to act trivially
on $\lambda$ and $\lambdabar$ (i.e., $\delta_{\cA,\cAbar}\lambda
= 0$ and $\delta_{\cA,\cAbar}\lambdabar = 0$); inside ``$[\quad]$''
on the right hand side, we first take the projection with respect
to powers of $p$, then reexpand the results into powers of $\cL$
and $\cLbar$ instead of $p$, and finally replace them by
$\lambda$ and $\lambdabar$.

\section{Riemann-Hilbert problem \label{sec:RHproblem}}

We are now in a position to apply the general machinery of
the preceding sections to two-dimensional string theory.
In this section, we solve a Riemann-Hilbert problem to
construct a special solution of the dispersionless Toda
hierarchy. In the next section, we prove that it indeed
describes the tachyon dynamics at self-dual radius by
showing that its $F$ potential satisfies
$w_{1+\infty}$-constraints of tachyon correlation
functions.

In general, Riemann-Hilbert problems for solving the
dispersionless Toda hierarchy can be written
\beqn
    \cLbar = f(\cL,\cM), \quad
    \cMbar = g(\cL,\cM),         \label{eq:generalRH}
\eeqn
where $\cL$, $\cM$, $\cLbar$ and $\cMbar$ are required
to be Laurent series of $p$ of the form assumed in
(\ref{eq:L},\ref{eq:M}); $f = f(\lambda,\mu)$ and
$g = g(\lambda,\mu)$ (``Riemann-Hilbert data'') are
functions satisfying the area-preserving condition
\beqn
    \lambda \frac{\rd f}{\rd \lambda} \frac{\rd g}{\rd \mu}
  - \frac{\rd f}{\rd \mu} \lambda \frac{\rd g}{\rd \lambda} = f
\eeqn
(which means that the map $(\log\lambda,\mu) \to (\log f, g)$
is area-preserving in the ordinary sense)
and some additional condition on its analyticity.
A general theorem \cite{bib:TT-dToda} ensures that
if (\ref{eq:generalRH}) has a unique solution, then
$\cL$, $\cM$, $\cLbar$ and $\cMbar$ satisfy all relevant
equations (\ref{eq:LaxL},\ref{eq:LaxM},\ref{eq:canonical})
of the Lax formalism.
Theoretically, one can thus obtain all solutions of the
dispersionless Toda hierarchy. Practically, explicit
solutions of such a Riemann-Hilbert problem is rarely
available.  Note that (\ref{eq:generalRH}) is just a
compact expression of an infinite number of highly
nonlinear relations between the two sets of variables
$(u_n,v_n)$ and $(\ubar_n,\vbar_n)$ (in which $t$,
$\tbar$ and $s$ enter as parameters); solving these
equations looks as difficult as solving the hierarchy
directly!  Fortunately, the Riemann-Hilbert problem
we consider below, is relatively easy to handle with.

The Riemann-Hilbert problem to be considered is the following:
\beqn
    \cL = \cMbar\cLbar, \quad
    \cLbar^{-1} = \cM \cL^{-1}.
                                            \label{eq:RH}
\eeqn
Apparently this does not take the form of (\ref{eq:generalRH}),
but can be readily rewritten in that form.  This non-standard
(but more symmetric) expression is rather suited for recognizing
a wedge algebra structure. The area-preserving condition, too,
can be easily checked. This Riemann-Hilbert problem can be
solved by almost the same method as used for the $A_{k+1}$
models \cite{bib:TT-dKP}. Actually, details of calculations
are rather similar to the case of the $D_\ell$ models
\cite{bib:T-DLG}; the integrable hierarchy underlying these
models, too, has four Lax functions, and Riemann-Hilbert
problems takes the same form as (\ref{eq:generalRH}).

Solving (\ref{eq:RH}) consists of several steps. The first
step is to split each equation of (\ref{eq:RH}) into two
pieces by applying $(\quad)_{\ge 0}$ and $(\quad)_{\le -1}$.
This gives the following four equations:
\beqnarray
    (\cL)_{\ge 0} &=&
       - \sum_{k=2}^\infty k \tbar_k (\cLbar^{-k+1})_{\ge 0}
       - \tbar_1 + s \cLbar
       + \sum_{n=1}^\infty \vbar_n \cLbar^{n+1},
                                           \label{eq:1a} \\
    (\cL)_{\le -1} &=&
       - \sum_{k=2}^\infty k \tbar_k (\cLbar^{-k+1})_{\le -1},
                                           \label{eq:1b} \\
    (\cLbar^{-1})_{\ge 0} &=&
       \sum_{k=2}^\infty k t_k (\cL^{k-1})_{\ge 0}
       + t_1,
                                           \label{eq:2a} \\
    (\cLbar^{-1})_{\le -1} &=&
         \sum_{k=2}^\infty k t_k (\cL^{k-1})_{\le -1}
       + s \cL^{-1}
       + \sum_{n=1}^\infty v_n \cL^{-n-1}.
                                           \label{eq:2b}
\eeqnarray

The second step is to decompose each equation into an infinite
number of equations not including $p$, by taking residue pairing
of both hand sides with suitable 1-forms. For instance, by taking
the residue pairing of both hand sides of (\ref{eq:1a},\ref{eq:1b})
with
(i) $\ubar_0 p^{-1} d\log p$,
(ii) $p^{n-1}d\log p$,
(iii) $\cLbar^{-n-1} d\log\cLbar$, respectively,
we can obtain the equations
\beqnarray
  \ubar_0
    &=& - \sum_{k=2}^\infty
            k\tbar_k \ubar_0
              \res[ \cLbar^{-k+1}p^{-1}d\log p ]
        + s,
                                             \label{eq:3'} \\
  u_n
    &=& - n\tbar_n \ubar_0{}^{n-1}
        - \sum_{k=n+1}^\infty
            k\tbar_k
              \res[ \cLbar^{-k+1} p^{n-1} d\log p ],
                                             \label{eq:45} \\
  \vbar_n
    &=&   \res[ (\cL)_{\ge 0} \cLbar^{-n-1} d\log\cLbar ]
                                                 \nonumber \\
    &&  + \sum_{k=2}^\infty
            k\tbar_k
                \res[ (\cLbar^{-k+1})_{\ge 0}
                    \cLbar^{-n-1}d\log\cLbar ]
                                             \label{eq:8}
\eeqnarray
for $n = 1,2,\ldots$. Here trivial equations of the form
$0 = 0$ have been omitted. It should be noted that this
process is reversible, because the 1-forms (i)-(iii) used
in the residue pairing form a complete set. Similarly, from
(\ref{eq:2a},\ref{eq:2b}), we obtain another infinite
set of equations
\beqnarray
  \ubar_0
    &=& \sum_{k=2}^\infty
          kt_k \res[\cL^{k-1} d\log p]
        + s                                   \label{eq:3} \\
  \ubar_n
    &=&   nt_n
        + \sum_{k=n+1}^\infty
            kt_k
              \res[ \cL^{k-1} p^{-n+1} d\log p ],
                                             \label{eq:67} \\
  v_n
    &=&   \res[ (\cLbar^{-1})_{\le -1} \cL^{n+1} d\log\cL ]
                                                \nonumber \\
    &&  - \sum_{k=2}^\infty
            kt_k
              \res[ (\cL^{k-1})_{\le -1} \cL^{n+1} d\log\cL ]
                                             \label{eq:9}
\eeqnarray
for $n = 1,2,\ldots$. This process, too, is reversible.
Therefore we now have only to solve these equations for
$u_n$, $v_n$, $\ubar_n$ and $\vbar_n$.

The third and final step is to solve these equations by Taylor
expansion.  Eqs. (\ref{eq:3'},\ref{eq:45},\ref{eq:3},\ref{eq:67})
include only $u$'s and $\ubar$'s.  By expanding these unknown
functions into Taylor series of $(t,\tbar)$ at $(t,\tbar)=(0,0)$,
one can convert these equations into (very complicated) recursion
relations of Taylor coefficients.  By the standard power
counting method, one can show that these recursion relations
uniquely determines $u$'s and $\ubar$'s as:
\beqnarray
    \ubar_0 &=& s + \ \mbox{higher order terms},
                                                  \nonumber\\
    u_n &=& - n \tbar_n s^{n-1} + \ \mbox{higher order terms },
                                                  \nonumber\\
    \ubar_n &=& n t_n + \ \mbox{higher order terms}
                                           \quad (n \ge 1).
\eeqnarray
Once $u$'s and $\ubar$'s are thus determined, remaining two
equations (\ref{eq:8},\ref{eq:9}) give $v_n$ and $\vbar_n$
explicitly. Thus our Riemann-Hilbert problem turns out to
have a unique solution.

The solutions $u_n$, $\ubar_n$, $v_n$ and $\vbar_n$ of the above
equations turn out to have good scaling properties. Note that
each equation of (\ref{eq:RH}) is invariant under the following
formal rescaling of variables included therein:
\beqnarray
   t_n \to c^{-n}t_n, && \tbar_n \to c^n \tbar_n,    \nonumber\\
             s \to s, && p \to c^{-1} p              \nonumber\\
     u_n \to c^n u_n, && \ubar_n \to c^{-n} \ubar_n, \nonumber\\
     v_n \to c^n v_n, && \vbar_n \to c^{-n} \vbar_n.
\eeqnarray
Since the Riemann-Hilbert problem has a unique solution, this means
that $u_n$, $\ubar_n$, $v_n$ and $\vbar_n$ indeed have the above
scaling property as functions of $(t,\tbar,s)$. In other words,
if we define a weight (U(1)-charge) of $t,\tbar,s$ as
\beqn
    \mbox{wt}(t_n) = -n, \quad
    \mbox{wt}(\tbar_n) = n, \quad
    \mbox{wt}(s) = 0,
\eeqn
then $u_n$, $\ubar_n$, $v_n$ and $\vbar_n$ become
quasi-homogeneous functions of degree $n$, $-n$, $n$ and $-n$,
respectively.  Accordingly, the functions $\phi$ and $F$,
which are defined by (\ref{eq:dphi},\ref{eq:dF}), become
quasi-homogeneous function of degree $0$.

Three remarks are now in order:

First, we have in fact two equations
(\ref{eq:3'}) and (\ref{eq:3}) that include $\ubar_0$ as a
main term; apparently this is redundant. Actually, one may
select one of them arbitrarily, and solve them along with
(\ref{eq:45},\ref{eq:67}). This eventually leads to the
same result, as one can verify by returning to
(\ref{eq:1a},\ref{eq:1b},\ref{eq:2a},\ref{eq:2b})
and reexamining the derivation of the above equations
therefrom.

Second, in the final step of the above
consideration, we have Taylor-expanded all unknown functions
at $(t,\tbar)=(0,0)$, but $s$ is left free. Namely, we do
not need Taylor expansion in $s$, and can set it to any
constant value. This is also reflected to the fact that
the weight (U(1)-charge) of $s$ is zero.  This is a
desirable property, because $s$ is interpreted to be
the cosmological constant of two-dimensional strings,
and an advantage of the Landau-Ginzburg formulation
lies in the fact that it describes the theoy with
non-zero cosmological constant.

Third, we have not specificed any explicit expression
of $u_n$, $v_n$, $\ubar_n$ and $\vbar_n$; they should
be very complicated, and we actually do not need
such explicit formulas. We just have to prove that
the Riemann-Hilbert problem has a unique solution.
The general machinery of the dispersionless Toda
hierarchy can work only after this fact is confirmed.
Once the existence of such a solution is proven,
one can derive $w_{1+\infty}$-constraints to the
$F$ potential therefrom, and identify it with the
generating function of tachyon correlation functions,
as we shall show in the next section. All relevant
information on the tachyon dynamics is now encoded
into the $F$ potential.

\section{Constraints to $F$ potential \label{sec:constraints}}

Let us now derive $w_{1+\infty}$-constraints to $F$,
To this end, we start from the relations
\beqn
    \cL^n = \cMbar^n \cLbar^n, \quad
    \cLbar^{-n} = \cM^n \cL^{-n}, \quad
    n = 1,2, \ldots,                          \label{eq:RH(n)}
\eeqn
which are an obvious consequence of (\ref{eq:RH}).  Just as
we derived (\ref{eq:3'}) etc. in the previous section, we now
take residue paring of both hand sides of (\ref{eq:RH(n)}) with
$d\cB_m$, $d\cBbar_m$ and $d\log p$ ($m = 1,2,\ldots$).
This results in the following relations:
\beqnarray
    \res[ \cL^n d\cB_m ]    &=& \res[ \cMbar^n \cLbar^n d\cB_m ],
                                                     \nonumber\\
    \res[ \cL^n d\cBbar_m ] &=& \res[ \cMbar^n \cLbar^n d\cBbar_m ],
                                                     \nonumber\\
    \res[ \cL^n d\log p]    &=& \res[ \cMbar^n \cLbar^n d\log p ],
                                                     \nonumber\\
    \res[ \cLbar^{-n} d\cB_m ]    &=& \res[ \cM^n \cL^{-n} d\cB_m ],
                                                     \nonumber\\
    \res[ \cLbar^{-n} d\cBbar_m ] &=& \res[ \cM^n \cL^{-n} d\cBbar_m ],
                                                     \nonumber\\
    \res[ \cLbar^{-n} d\log p]    &=& \res[ \cM^n \cL^{-n} d\log p ].
                                                 \label{eq:resRH(n)}
\eeqnarray
Note that these relations conversely imply (\ref{eq:RH(n)}),
because this residue pairing is complete (i.e.,
$\res[fd\cB_m] = \res[fd\cBbar_m] = \res[fd\log p] = 0$ for
all $m=1,2,\ldots$ if and only if $f = 0$).
We can now apply (\ref{eq:deltav}) to each equations of
(\ref{eq:resRH(n)}) to rewrite them as:
\beqnarray
   &&   \dfrac{\rd}{\rd t_m}\delta_{\cL^n,\cMbar^n \cLbar^n}F
      = \dfrac{\rd}{\rd \tbar_m}\delta_{\cL^n,\cMbar^n \cLbar^n}F
      = \dfrac{\rd}{\rd s}\delta_{\cL^n,\cMbar^n \cLbar^n}F
      = 0,
                                                        \nonumber\\
   &&   \dfrac{\rd}{\rd t_m}\delta_{\cM^n \cL^{-n},\cLbar^{-n}}F
      = \dfrac{\rd}{\rd \tbar_m}\delta_{\cM^n \cL^{-n},\cLbar^{-n}}F
      = \dfrac{\rd}{\rd s}\delta_{\cM^n \cL^{-n},\cLbar^{-n}}F
      = 0.
\eeqnarray
These equations show that $\delta_{\cL^n,\cMbar^n \cLbar^n}F$
and $\delta_{\cM^n \cL^{-n},\cLbar^{-n}}F$ are constant.
Actually, this constant should vanish: If one recalls the
aforementioned scaling properties of $v_n$, $\vbar_n$ and
$\phi$, and apply them to general formula (\ref{eq:deltaF}),
one will be able to see that $\delta_{\cL^n,\cMbar^n \cLbar^n}F$
and $\delta_{\cM^n \cL^{-n},\cLbar^{-n}}F$ are quasi-homogeneous
of degree $-1$. This means that the constant values should be
zero. Thus we can conclude that $F$ satisfies the equations
\beqn
    \delta_{\cL^n,\cMbar^n \cLbar^n}F = 0,    \quad
    \delta_{\cM^n \cL^{-n},\cLbar^{-n}}F = 0, \quad
    n = 1,2,\ldots.
                                     \label{eq:w-constraints}
\eeqn
Furthermore, by carefully examining the above derivation, one
can see that this derivation is reversible; Eqs. (\ref{eq:RH(n)})
(therefore the original Riemann-Hilbert problem) can be derived
conversely from (\ref{eq:w-constraints}).

Eqs. (\ref{eq:w-constraints}) are, actually, just a disguise
of the $w_{1+\infty}$-constraints of Hanany et al.  By general
formula (\ref{eq:deltaF}), one can rewrite (\ref{eq:w-constraints})
into a more explicit form:
\beqnarray
    v_n - \frac{1}{n+1} \res[ \cMbar^{n+1}\cLbar^n d\log\cLbar]
      &=& 0,                                    \nonumber\\
    \vbar_n + \frac{1}{n+1} \res[ \cM^{n+1} \cL^{-n} d\log\cL]
      &=& 0.
\eeqnarray
One can then substitute $v_n = \rd F/\rd t_n$ and $\vbar_n =
- \rd F/\rd \tbar_n$ to write the left hand side in terms
of derivatives of $F$.  Furthermore, one can introduce a
new variable $X$ and, as in (\ref{eq:L->lambda}),
rewrite the residues in terms of $X$ by replacing
$\cL \to X$, $\cLbar \to X^{-1}$.  Thus, eventually,
(\ref{eq:w-constraints}) turn into the following form:
\beqnarray
    \dfrac{\rd F}{\rd t_n}
    - \frac{1}{n+1}
        \res \left[
          \left( \frac{\cMbar(\cLbar\to X^{-1})}{X}
                                                \right)^{n+1}
             dX \right]
                &=& 0,                           \nonumber \\
    \dfrac{\rd F}{\rd \tbar_n}
    + \frac{1}{n+1}
        \res \left[
           \left( \frac{\cM(\cL\to X)}{X} \right)^{n+1}
              dX \right]
                &=& 0,
                                         \label{eq:HOPconstraints}
\eeqnarray
which become exactly the $w_{1+\infty}$-constraints of Hanany et al.
if we interpret their two Landau-Ginzburg potentials $W(X)$,
$\Wbar(X)$ and tachyon correlation functions $<\!< T_n >\!>$ as:
\beqnarray
    W(X) = - \frac{\cM(\cL\to X)}{X}, &&
    \Wbar(X) = - \frac{\cMbar(\cLbar\to X^{-1})}{X},
                                          \label{eq:LGpotential} \\
    \dbra T_n \dket = \frac{1}{n} \dfrac{\rd F}{\rd t_n}, &&
    \dbra T_0 \dket = \dfrac{\rd F}{\rd s},             \nonumber\\
    \dbra T_{-n} \dket = - \frac{1}{n} \dfrac{\rd R}{\rd\tbar_n},
                                          && (n = 1,2,\ldots).
                                          \label{eq:Tcorrelator}
\eeqnarray
The extra numerical factors on the right hand side emerge
because our $(t,\tbar,s)$ are slightly different from the
background sources of Hanany et al.  Our results agree with
theirs if we interpret the correlator $<\!< \cO >\!>$ as:
\beqn
    \dbra \cO \dket
    = \left<   \cO \exp(
               \sum_{n=1}^\infty n t_n T_n
               + s T_0
               - \sum_{n=1}^\infty n \tbar_n T_{-n} )
      \right>.
\eeqn

Actually, in place of (\ref{eq:RH(n)}), one can consider even
more general combinations of the fundamental Riemann-Hilbert
relation as:
\beqn
    \cM^k \cL^{n-k} = \cMbar^n \cLbar^{n-k}, \quad
    k,n = 0,1,\ldots.               \label{eq:RH(kn)}
\eeqn
Then, by the same reasoning as above, the following constraints
can be obtained:
\beqn
    \delta_{\cM^k \cL^{n-k},\cMbar^n \cLbar^{n-k}} F = 0.
                                    \label{eq:constraints(kn)}
\eeqn
In terms of the Landau-Ginzburg potential, more explicitly,
these constraints can be written
\beqn
    \frac{1}{k+1}
      \res \left[ \left( - W(X) \right)^{k+1} X^n dX \right]
  = \frac{1}{n+1}
      \res \left[ \left( - \Wbar(X) \right)^{n+1} X^k dX \right].
\eeqn
Of course, as also noted by Hanany et al., their
$w_{1+\infty}$-constraints are in themselves powerful
enough to determine the tachyon correlation functions
completely. In this respect, the above constraints are
redundant. These extra constraints, however, turn out
to stem from underlying highersymmetries, as we shall
discuss in the next section.

\section{States and fields generated by wedge algebra
\label{sec:wedge}}

We first note that both hand sides of (\ref{eq:RH(kn)})
are generators of a wedge algebra.  To clarify this fact,
we introduce nonnegative half-integer indices $(j,m)$
in the ``wedge'' $|m| \le j$ by the usual convention
\beqn
    k = j-m, \quad n = j+m,
\eeqn
and write both hand sides of (\ref{eq:RH(kn)}) as $w_{jm}$:
\beqn
     w_{jm} = \cL^n (\cM\cL^{-1})^k
            = (\cMbar\cLbar)^n \cLbar^{-k}.
\eeqn
Since
\beqn
    \{ \cL, \cM\cL^{-1} \}
    = \{ \cMbar\cLbar, \cLbar^{-1} \} = 1,
\eeqn
$w_{jm}$ indeed form a wedge algebra with respect to the Poisson
bracket.  In the following, we propose a speculative interpretation
of this wedge algebra as generators of ``extra'' states and fields
of two-dimensional strings.

Let us show how such ``states'' emerge in our framework.
Let $W_{jm}$ denote the following symmetries of the
dispersionless Toda hierarchy:
\beqn
    W_{jm} = \delta_{\cL^n (\cM\cL^{-1})^k,0}
           = - \delta_{0,(\cMbar\cLbar)^n \cLbar^{-k}}.
\eeqn
These symmetries are understood to be acting on the ring
$\cR$ of Section \ref{sec:symmetries}. The two expressions
on the right hand side give the same symmetry because of
(\ref{eq:constraints(kn)}).  Furthermore, by (\ref{eq:CR}),
$W_{jm}$ obey the same commutation relations as the Poisson
commutation relations of $w_{jm}$; the central terms disappear,
as usual, on a wedge.  The action of those sitting on the
``edge" of the wedge, $(j,m) = (n/2,\pm n/2)$ generate the
tachyon correlation functions:
\beqnarray
  && W_{n/2,n/2} F  = - \frac{\rd F}{\rd t_n}
                    = - n \dbra T_n \dket,          \nonumber\\
  && W_{n/2,-n/2} F =   \frac{\rd F}{\rd \tbar_n}
                    = - n \dbra T_{-n} \dket.
\eeqnarray
In view of this, we propose to consider the action of other
$W$'s, too, as insertion of a ``state" $W_{jm}$ into the
correlator:
\beqn
    W_{j_1,m_1}\cdots W_{j_r,m_r} F
    = \dbra W_{j_1,m_1} \cdots W_{j_r,m_r} \dket.
\eeqn
Commutation relations (\ref{eq:CR}) of our symmetries will
then reproduce the $w_{1+\infty}$ Ward identities in the
matrix model approach \cite{bib:matrix-model} (now in
in the presence of tachyon backgrounds).

What about ``fields"?   A set of fields $\phi_n(X)$ and
$\phibar_n(X)$ are introduced by Hanany et al. \cite{bib:HOP}
as $c=1$ analogues of $c<1$ chiral ring generators and
gravitational descendents. In our interpretation of $(t,\tbar)$
as background sources, $\phi_n(X)$ are given by
\beqn
   \phi_n(X) = - \frac{1}{n}\frac{\rd W(X)}{\rd t_n}, \quad
   \phi_{-n}(X) = \frac{1}{n}\frac{\rd W(X)}{\rd \tbar_n} \quad
   (n = 1,2,\ldots),
\eeqn
and $\phibar_n(X)$ by similar derivatives
of $\Wbar(X)$.  Since the Landau-Ginzburg potentials are written
in terms of $\cM(\cL\to X)$ and $\cMbar(\cLbar\to X^{-1})$ as
shown in (\ref{eq:LGpotential}), these ``fields'' are exactly
the same quantities as emerging on the right hand side of
(\ref{eq:dB(lambda)dM(lambda)}), i.e., derivatives of the flow
generators $\cB_n$ and $\cBbar_n$ with respect to the
Landau-Ginzburg field variable $X$. Note that this is parallel
to the construction of chiral ring generators in the $A_{k+1}$
models \cite{bib:DVV,bib:Krichever,bib:Dubrovin}. These ``fields''
are Landau-Ginzburg counterparts of tachyon ``states''
$W_{n/2,\pm n/2}$.  To find other ``fields'', let us note that
$\phi_n(X)$ can also be written
\beqn
    \phi_n(X) = X^{n-1} + \frac{1}{n}\delta_{\cL^n,0} W(X),  \quad
    \phi_{-n}(X) = - \frac{1}{n}\delta_{0,\cLbar^{-n}} W(X).
\eeqn
Here we have used (\ref{eq:deltaM(lambda)}), recalling the
correspondence (\ref{eq:LGpotential}) between the Laudau-Ginzburg
potential and the Lax functions. The somewhat strange extra term
$X^{n-1}$ is due to the presence of tachyon backgrounds. Since the
symmetries on the right hand side are just $W_{n/2,\pm n/2}$,
we are naturally led to conjecture that
``fields'' $\Phi_{jm}(X)$ corresponding to the ``states''
$W_{jm}$ are to be given by
\beqn
    \Phi_{jm}(X) = W_{jm} W(X).
\eeqn
Similarly the action of $W_{jm}$ on $\Wbar(X)$ will give another
set of extra ``fields'' $\Phibar_{jm}(X)$.  In principle, one
can find an explicit form of these ``extra fields'' from
(\ref{eq:deltaM(lambda)}), though it will become considerably
complicated in general. To push forward this speculation further,
we will have to examine if the period integral representation
of tachyon correlation functions and the contact algebra of
$\phi_n(X)$ and $\phibar_n(X)$
\cite{bib:Ghoshal-Mukhi,bib:HOP,bib:Danielsson} can be extended
to our $\Phi_{jm}(X)$ and $\Phibar_{jm}(X)$.

\section{Conclusion \label{sec:conclusion}}

Inspired by the suggestion of Hanany et al., we have considered
the integrable structure of two-dimensional string theory at
self-dual compactification radius. Our main conclusion is that
the dispersionless Toda hierarchy is a very convenient framework
for studying the tachyon sector of this theory. We have been
able to identify a special solution of this integrable hierarchy
in which full data of the tachyon dynamics is encoded.
The $w_{1+\infty}$-constraints of tachyon correlation functions
can be indeed reproduced from the construction (Riemann-Hilbert
problem) of this solution. The Landau-Ginzburg formulation, too,
turns out to be closely related to the Lax formalism of the
dispersionless Toda hierarchy. Furthermore, we have pointed out
the existence of a wedge algebra structure behind the
Riemann-Hilbert problem, and proposed a speculative
interpretation of this algebra as generators of ``extra''
states and fields in this model of two-dimensional strings.
The last issue deserves to be pursued in more detail.

We conclude this paper with several remarks.

1) In the context of two-dimensional gravity, the dispersionless
Toda hierarchy is zero-genus limit of the full Toda hierarchy.
A full-genus analysis in the language of the Toda hierarchy is
done by Dijkgraaf et al. \cite{bib:DMP}.  We will be able to
extend the results of this paper to that case.

2) As already mentioned, the integrable hierarchy underlying
the topological $D_\ell$ models \cite{bib:T-DLG} resembles the
dispersionless Toda hierarchy. This hierarchy is related to the
Drinfeld-Sokolov hierarchy of $D$-type. It is intriguing that
Danielsson \cite{bib:Danielsson} pointed out a link between a
deformed Landau-Ginzburg model and the Drinfeld-Sokolov hierarchy
of $D$-type.

3) Our method for solving a Riemann-Hilbert problem can be extended
to more general cases such as:
\beqn
    \cL^N  = \cMbar \cLbar^{\Nbar} / \Nbar, \quad
    \cLbar^{-\Nbar} = \cM \cL^{-N} / N,
                                        \label{eq:NNbar}
\eeqn
where $N$ and $\Nbar$ are nonzero integers. In this paper,
we have considered the simplest case, $N=\Nbar=1$; other cases,
too, may have interesting physical interpretations.
For instance, the work of Dijkgraaf et al. \cite{bib:DMP}
implicitly shows that if the compactification radius
($\beta$ in their notation) is a positive integer, the dynamics
of tachyons in zero-genus limit can be described by the solution
of (\ref{eq:NNbar}) with $N=\Nbar=\beta$. Thus we can deal with
a discrete series of theories at non-self-dual ($\beta > 1$)
radii in much the same way; a full genus analysis will become
possible in the full Toda hierarchy.

4) Discrete states and quadratic Ward identities in the free field
approach \cite{bib:free-field} are still beyond our scope. Our
approach by the dispersionless Toda hierarchy is at most an
effective theory in the tachyon sector, though we can anyhow
reproduce the wedge algebra symmetries acting on tachyon states.
Presumably, a suitable integrable extension of the dispersionless
(or full) Toda hierarchy will provide a framework for dealing
with this issue.

\begin{acknowledgement}
The author is very grateful to Hiroaki Kanno and
Takashi Takebe for many useful comments. This work
is partially supported by the Grant-in-Aid for
Scientific Research, the Ministry of Education,
Science and Culture, Japan.
\end{acknowledgement}

\begin{note-added}
Hiroaki Kanno informed the author that Tohru Eguchi
independently arrived at the same Riemann-Hilbert
relation as ours (\ref{eq:RH}).
\end{note-added}

\end{document}